\documentclass[11pt,a4paper]{article}

\usepackage{textcomp}
\usepackage{tikz}
\usepackage{amsmath,amssymb,amsthm,appendix,placeins, floatrow}
\usepackage[utf8]{inputenc}
\usepackage{setspace}
\usepackage{changepage}
\usepackage{booktabs,threeparttable,dcolumn,tabularx,longtable,rotating}
\usepackage{subcaption}
\usepackage[hidelinks=true]{hyperref}
\usepackage{natbib}
\usepackage{lscape}	
\usepackage{chngcntr}


\newcommand{\sym}[1]{\ensuremath{^{#1}}}

\begin{document}
\onehalfspace

\title{\Large{Labor Market Integration of Refugees: RCT Evidence from an Early Intervention Program in Sweden\footnote{We thank Mette Foged, Matti Sarvim\"{a}ki, Johan Vikstr\"{o}m, and seminar participants at the EALE SOLE AASLE World Conference 2020, at the 2022 American Economic Association Meeting, at the Department of Statistics at Uppsala University, at IFAU, at the UIL Workshop in Uppsala, and at Helsinki GSE Labor and Public economics seminar for valuable comments and suggestions. We also thank  Ann-Sofie Arvidsson and Asllan Hajdari for helping out implementing the randomization, as well as S\"{u}kran Dogan, Peter de Jonge and Ulf Wallin for sharing valuable insights about the program in Gothenburg and for organizing visits on location in Gothenburg. Johan Egebark acknowledges financial support from Jan Wallanders och Tom Hedelius stiftelse. AEA RCT ID: AEARCTR-0004788. This paper uses confidential data from the Swedish PES and the City of Gothenburg. The data can be obtained by filing a request directly with IFAU (ifau@ifau.uu.se). The authors are willing to assist (Ulrika Vikman, Ulrika.vikman@ifau.uu.se). Do files are available in the Online Appendix.}}} 

\author{Matz Dahlberg\footnote{Uppsala University, Uppsala, Sweden; CESifo; IFAU.} \and Johan Egebark\footnote{Swedish Public Employment Service, Stockholm, Sweden.} \and Ulrika Vikman\footnote{The Institute for Evaluation of Labour Market and Education Policy (IFAU), Uppsala, Sweden. Corresponding author, Ulrika.vikman@ifau.uu.se} \and G\"{u}lay \"{O}zcan\footnote{Swedish Public Employment Service, Stockholm, Sweden.} }
%\date{June, 2023}
\maketitle

\vspace*{-1.0cm}

\begin{abstract}
\footnotesize{This study uses a randomized controlled trial to evaluate a new program for increased labor market integration of refugees. The new program starts shortly after the residence permit is granted and uses three main components: early and intensive language training, work practice with supervisors, and job search assistance performed by professional caseworkers. The immediate and intensive assistance contrasts previous integration policies which typically constitute low-intensive help over long periods of time. We find large positive effects on employment of the program, with magnitudes corresponding to around 15 percentage points. A mediation analysis shows that 7--8 percent of the impact of the program is explained by increases in documented language skills. Cost estimates suggest that the new policy is less expensive than earlier labor market programs.} 
\end{abstract}

\thispagestyle{empty}
\clearpage

%%%%%%%%%%%%%%%%%%%%%%%%%%%%%%%%%%%%%%%%%%%%%%%%%%%%%%%%%%%%%%%%%%%%%%%%%%%%%%%%%%%%
\section{Introduction} \label{sec:intro}
%%%%%%%%%%%%%%%%%%%%%%%%%%%%%%%%%%%%%%%%%%%%%%%%%%%%%%%%%%%%%%%%%%%%%%%%%%%%%%%%%%%%
It is well-established that labor market integration takes time (\citealp{cortes_are_2004}; \citealp{ruiz_differences_2018}; \citealp{marten_ethnic_2019}; \citealp{brell_labor_2020}; \citealp{fasani_lift_2021}). In the OECD, employment among new immigrants is ten percentage points lower than among the native-born (\citealp{oecdeu_settling_2018}), and, across Europe, only 62\% of refugee men and 45\% of refugee women are employed (\citealp{oecdec_how_2016}). In recent years, large inflows of refugees to many Western countries have spurred a new, and sometimes intense, debate on how to best integrate immigrants in their new countries. Since getting immigrants well-integrated is considered crucial, from the perspective of both the individuals and the society, finding successful integration policies is at the top of the political agenda (\citealp{dustmann_economics_2017}; \citealp{HangartnerSarvimaki}; \citealp{battisti_can_2019}).

Previous research shows that the initial conditions that immigrants face in their host countries have important and long-lasting impacts on their future economic and social integration (\citealp{aslund_when_2007}; \citealp{braun_local_2017}; \citealp{bansak_improving_2018}; \citealp{bratu_age_2021}; \citealp{fasani_lift_2021};\citealp{aksoy_first_2020}). Given the importance of initial conditions, it is striking that help to recent immigrants -- such as language courses, work practice, and job search programs -- is typically offered in small doses and over long periods of time (\citealp{andersson_joona_2020, hangartner_managing_2021, arendt_refugee_2022}). In contrast, an immediate injection of highly intensive assistance is potentially more gainful if the goal is to achieve fast integration. In this study, we use a randomized controlled trial (RCT) to evaluate a new labor market program that uses an early intervention strategy. 

The evaluation is based on a field experiment where 140 newly arrived refugees in the city of Gothenburg in Sweden were randomly assigned to either a treatment (50 percent) or a control group. While the individuals in the treatment group were invited to participate in the program, the control group joined the baseline services at the Swedish Public Employment Service (PES). The program contains three main components: early and intensive language training, intensive work practice with supervisors, and job search assistance performed by professional caseworkers. The language training starts immediately after the refugees enroll in the program, and lasts for three months. This is followed by six months of supervised work practice, where the participants are trained within a specific occupation. Finally, after the work practice ends, the participants are offered help from professional caseworkers with finding a job. 

If we compare the program we evaluate to other integration policies studied before, both in Sweden and in other countries, three distinct features stand out. First, by combining various types of active labor market policies within a short period of time, the program has a comparably high treatment intensity. Our comparisons show substantial differences in intensity between treated and non-treated individuals throughout the entire period that the program runs. As an example, the refugees in the treatment group get on average around 160 hours \textit{more} language training than those in the control group. Second, and importantly, the program introduces the intensive assistance shortly after the residence permit is granted. The gain from immediate assistance is potentially very large, since most of the employment growth takes place in the early years after a refugee gets a residence permit. One of the goals of the new program was to increase the likelihood of early transitions to employment, to avoid additional losses of human capital and decreased motivation due to long waiting times.\footnote{Recent evidence suggests that long waiting times have significant and lasting negative effects on refugees' employment growth (\citealp{hainmueller_when_2016, hvidtfeldt_estimate_2018, marbach_long-term_2018,fasani_lift_2021}). } Third, the program targets a sub-group of immigrants that is increasingly relevant for policy makers: newly arrived refugees with low education. Low-skilled refugees comprises a large and important part of the influx of immigrants to Western countries in the past years, and many receiving countries struggle to find ways of integrating this group due to its weak initial position in the labor market. 90 percent of the individuals in our sample have no formal education above elementary school (from the home country), and the average number of days since the residence permit was granted is 220--230. This confirms that the targeted refugees have almost no formal education and have just started their integration process when they enter the program.

We present compelling evidence of the integration program having positive effects on employment. Around 30\% of the individuals in the treatment group are employed each month during the first year after the end of the program, compared to an average of approximately 15\% in the control group. The results are insensitive to a number of robustness checks, including using p-values from Fisher's exact testing instead of relying on asymptotic properties for inference. Our findings in this paper thus strongly support the idea that early and intensive interventions are successful in getting newly arrived, low-educated refugees integrated into the labor market. 

Recent research on immigrant integration has a particular interest in the importance of language proficiency (\citealp{lochmann_effect_2019, arendt_trade-offs_2023, Foged_Werf}). Since we have access to detailed data on documented language skills, we can measure the importance of increased language proficiency as a mediator for the estimated employment effects. We first document a significant positive impact of the program on documented skills. At the end of the program, the treated refugees are almost 30 percent more likely to have completed the courses within the Swedish language training. This finding is in line with results in \cite{Foged_Werf} who find positive effects on language skills of extra language training in Denmark. Next, we follow \cite{Heckman2013} and present results showing that increased host-country language skills account for almost 10 percent of the effect on employment. 

The integration program that we consider was designed as a direct reaction to the poor labor market situation for newly arrived refugees in the Gothenburg region. One relevant question is to what extent the situation in Gothenburg can be generalized to other settings. By compiling data for the full population of low-educated newly arrived refugees in Sweden for the years 2017--19, we show that refugees in other regions, including the other two major Swedish cities Malm\"{o} and Stockholm, have very similar employment trends as our control group. Irrespective of which region we consider, employment for non-treated low-educated refugees grows slowly, from zero to roughly 15--20 percent during the first 2--2,5 years. In addition, our control group in Gothenburg displays similar employment trends as refugees in other OECD countries, such as the other Nordic countries, Germany, and Australia (see Figure 2 in \citealp{brell_labor_2020} and Figure 3 in \citealp{Foged_Hasager_Peri_NBER}). This suggest that the program would benefit refugees in other regions in Sweden, and potentially in other countries as well.

Finally, we analyze to what extent the program is expensive. We calculate that the total cost of the program amounts to around SEK 1.2 million (EUR 126,000), corresponding to SEK 27,000 per participant.\footnote{In April 2017, 1 Euro equals 9.5 SEK and 1 US Dollar equals 8.9 SEK.} Comparing this cost to the cost of other labor market programs in Sweden it is clear that our program is relatively inexpensive. 

The rest of the paper is organized as follows. While the next section describes the integration program and presents the experimental design, section 3 provides the details on the estimation method, the data and show characteristics of our population. Section 4 presents the results, section 5 discusses the external validity of our results, and, finally, section 6 concludes by discussing the results.

\section{The early intervention program and the RCT design}
Our study evaluates a new integration program that was developed and implemented in the city of Gothenburg in Sweden. The program came into effect as a result of a local initiative involving the Public Employment Service (PES), the City of Gothenburg, and the largest real estate company in Gothenburg (F\"{o}rvaltnings AB Framtiden).\footnote{The Swedish PES belongs to the central government sector. However, each municipality in Sweden, including Gothenburg, has its own autonomous local PES offices. The City of Gothenburg, which belongs to the local public sector, is responsible for providing education at different levels, including adult education. F\"{o}rvaltnings AB Framtiden, founded in 1915, is one of the largest real estate companies in Sweden. The company is owned by the municipality but its organizational structure corresponds to that of a private firm.} The goal of the new program, as stated by the three founders of the program, was to substantially improve the long-run labor market prospects of newly arrived immigrants through faster transitions to employment. Below we describe the program and the institutional setting in more detail, and present the way we designed the experiment.

\subsection{Three building blocks}\label{sec:3blocks}
The program includes three main components: (1) language training,  (2) supervised work practice, and (3) job search assistance. The language training starts immediately after the refugees enroll in the program, and lasts for three months. This is followed by six months of supervised work practice. After the nine months of language training and work practice, the participants are offered help from professional PES caseworkers with finding a job. All the activities included in the program are similar to those found in the baseline services at the Swedish PES. The main difference is that, in contrast to the baseline scenario, participants are exposed to an \textit{immediate intervention} consisting of \textit{highly intensive} assistance. Everyone who joins the program are expected to participate full time during the approximately ten months that the program runs.

\subsubsection*{Language training}
In Sweden, all newly arrived refugees who register as unemployed at the Swedish PES are offered Swedish language classes, so-called Swedish for Immigrants (SFI).\footnote{Swedish for Immigrants is provided free of charge by the local government sector. There are three different tracks: track 1, which is offered to illiterates, includes four levels of courses (A--D), track 2 includes three levels (B--D), and track 3, which is typically offered to those with some higher education, includes two levels (C and D).} This typically amounts to 15 hours of teaching per week throughout a two-year integration period (in addition to this, individuals are assumed to study and practice Swedish on their own). One of the aims of the new program is to provide more intensive language training at an early stage, and, therefore, all participants are obliged to attend additional hours of language classes during the first three months of the program. After three months, the participants continue with SFI at the same pace as the control group, i.e., two days (roughly 15 hours) per week.\footnote{During the work practice, in the second phase of the program, the participants are expected to continue SFI until they reach level D, which is the required level for being able to enter the regular Swedish education system.}

In addition to the initial SFI language training, the program includes other, more general courses, as well. This includes civic orientation classes which introduce the participants to the Swedish society\footnote{This is in addition to a 60-hour basic civic orientation class taken by everyone in the integration program. The classes concern practical matters, including how to apply for a rental apartment, how to pay bills, etc. The additional civic orientation offered to the treatment group adds more detailed knowledge.}, one course teaching workplace rules\footnote{This course informs the participants what is expected of them as employees, such as being on time and reporting sick leave to the employer. It also considers more general matters, such as how to get unemployment insurance in Sweden.}, one course teaching how to be service-minded, and introductory courses in mathematics and IT. All extra courses are given in Swedish and they run throughout the whole program period (in parallel with the other activities). In sum, during the first three months of the program, the SFI courses and the additional courses amounts to (close to) 40 hours of classroom training per week.

\subsubsection*{Work practice}
After the initial intensive language training, the participants continue to the next level of the program as they start work practice. The work practice proceeds three days a week for six months (with the remaining two days per week dedicated to SFI and the extra courses described above). Each participant is appointed 1--3 supervisors, who have the main responsibility for the intern. When allocating the participants to a workplace, two aspects are considered. First, in order to challenge the participants to practice Swedish, everyone is assigned supervisors who do not speak their native language. Second, to get the participants used to commuting, they are assigned a workplace outside their own residential area.

Work practice is a widely used labor market policy in Sweden. According to figures from the PES, 25--30,000 job seekers per year participate in this activity. Hence, this type of training is not unique to our program -- all job seekers who register with the PES, including our control group, have access to it. The PES cooperates with firms in the local labor market to find suitable workplaces, typically within industries where labor demand is high. This is also the case in the program we evaluate. 

\subsubsection*{Job search assistance}
Finally, as the work practice ends, efforts begin in terms of finding potential employers. The job search assistance in this stage is performed by professional PES caseworkers, and is significantly more intensive than the support offered to job seekers in the baseline services. In practice, the counselling is about helping the participants find suitable vacant jobs matching their level of skills, and helping with applications, CV-writing, interview training etc. The PES caseworkers also receive help from representatives from the real estate company, who contribute by providing insights on local labor demand.

\subsection{The experimental design} \label{experimental_design}
In total, four waves of refugees have started the program, one per year during 2016--20.\footnote{The first, second, third and fourth wave started in October 2016, May 2017, April 2018, and January 2020 respectively.} In each wave, there have been around 50 slots available. Our study follows the wave where treatment was randomized, the one starting in 2017. The eligibility criterion that was used in the program is comparably strict. First, only those with less than high school education (from the country of origin) were eligible, and, second, participation required recently having received a Swedish residence permit. Based on these criteria, the PES in Gothenburg identified 140 potential participants in April 2017.\footnote{Since the program includes work practice, individuals who were not ready to start employment in the short run, mainly due to different types of health problems, were excluded from the pool of potential candidates. Note that the 140 individuals constituted the universe of eligible individuals in Gothenburg at the time of the start of the program.}

Out of the 140 potential participants, 70 individuals were randomly drawn to a treatment group and 70 were randomly drawn to a control group. The 70 individuals in the treatment group got an offer to start the program, and the 70 individuals in the control group continued within the baseline services provided by the PES in Gothenburg. Since previous studies show that the effects of active labor market programs differ between men and women (\citealp{card_what_2018}), we block-randomized based on gender to get better precision. This makes the share of men and women in the treatment group identical to the shares in the target population.\footnote{Based on their shares in the target population, we block-randomized 27 (out of 53) women and 43 (of 87) men to the treatment group. The randomization was carried out by researchers at the Swedish PES in Stockholm.} 

\begin{figure}[htbp] \centering
\caption{Illustration of the different stages of the experiment.}\label{fig:illustration}
\begin{tikzpicture}

\draw[thick,| ->] (-2.5,0) --   (0,0) node[align=left, below] at (-1.3,0) 
{\scriptsize \textbf{April 2017} \\ 
\scriptsize Randomization, \\
\scriptsize summons, meeting\\
\scriptsize (Month 0)};

\draw[thick,| ->] (0,0) --   (2.5,0) node[align=left, below] at (1,0) 
{\scriptsize \textbf{May 2017} \\ 
\scriptsize Meeting, \\
\scriptsize preparation \\
\scriptsize (Month 1)};

\draw[thick,| ->] (2.5,0) --   (5,0) node[align=left, below] at (3.7,0) 
{\scriptsize \textbf{June--Aug. 17} \\
\scriptsize Intensive \\
\scriptsize language course \\ 
\scriptsize  (Months 2--4)};

\draw[thick,->] (5,0) --   (7.5,0) node[align=left, below] at (6.2,0) 
{\scriptsize \textbf{Sep. 17--Feb. 18} \\
\scriptsize Supervised \\
\scriptsize work practice \\ 
\scriptsize (Months 5--10)};

\draw[thick, ->] (7.5,0) --   (10,0) node[align=left, below] at (8.7,0) 
{\scriptsize \textbf{March 2018--} \\
\scriptsize Job search \\
\scriptsize assistance \\ 
\scriptsize (Months 11--) \bigskip};
\end{tikzpicture}

\floatfoot{\emph{Note}: The randomization was conducted on April 22, 2017. The first information meetings were held in the week following the randomization. The second information meeting, given to those who accepted the offer to participate, was held on May 22. The activities started on May 29, 2017. The job search assistance did, in some cases, start before March 2018. \\ }
\end{figure}
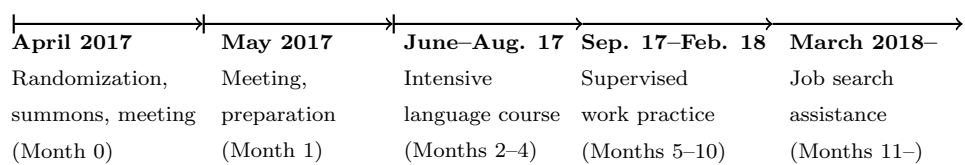

Figure \ref{fig:illustration} illustrates the different stages of the experiment. All 70 individuals in the treatment group were summoned to an information meeting, which took place at the PES in Gothenburg at the end of April 2017 (less than one week after the randomization). In the analysis, we denote April 2017 as month zero. All 70 individuals who were summoned showed up at the meeting, which means that everyone in the treatment group were given a detailed description of the program by the representatives from the three founders of the program. At the end of the information meeting everyone in the treatment group were offered to participate. 44 individuals accepted the offer and started the program. Those who accepted were summoned to a second information meeting, which was held at the end of May 2017, just before the program activities started. In the analysis, we denote May 2017 as the first month of the program, since the individuals in the treatment group may react already based on the information provided at the meetings. The individuals in the treatment group who turned down the offer participated in activities within the baseline services at the PES (which means they followed the same track as the control group).

The program activities started in late May 2017 when the participants began the intensive language training. This lasted for three months, months 2--4 after the randomization (June to August 2017), and was then followed by six months of supervised work practice, months 5--10 after randomization (from September 2017 to February 2018). Finally, in the last stage, the participants were offered job search assistance as they finished the work practice and were ready to take a job. The assistance typically started in March 2018, 11 months after the randomization.

\subsection{What the treatment and control individuals \textit{actually} do during the program period}
\begin{figure}[htbp] \centering
\caption{Timing and intensity of treatment.}\label{fig:treatment}
\includegraphics[width=1\textwidth]{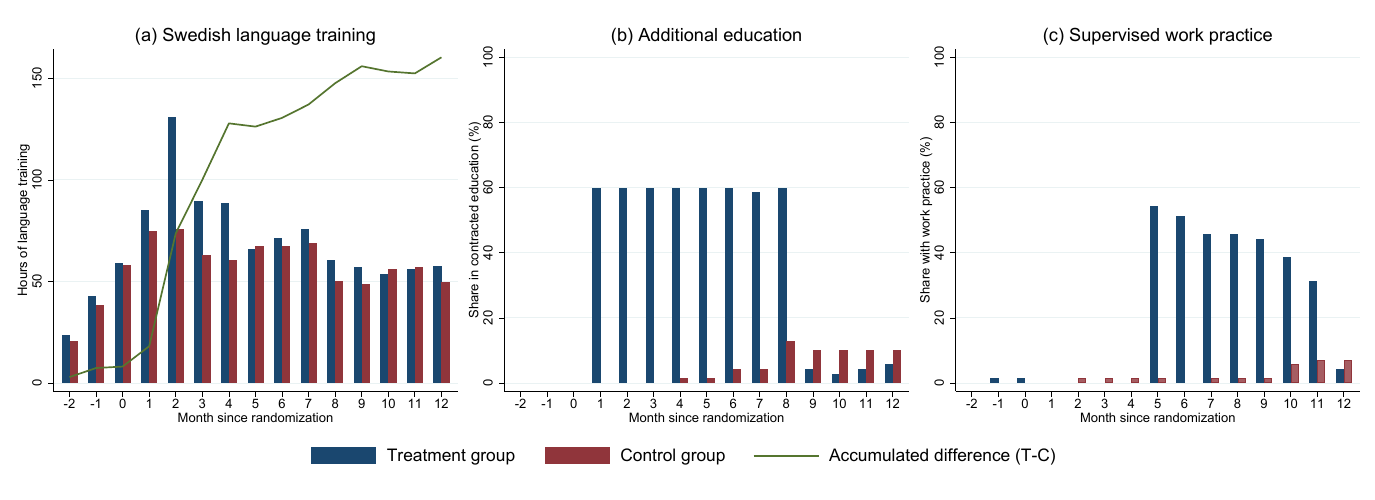} 
\floatfoot{\emph{Notes:} The figure compares treated and non-treated refugees in terms of activities. We follow both groups for 12 months after the start of the program, and for two months before the program starts. Hours of Swedish language training is the number of registered hours by the provider of the language courses. Additional education and supervised work practice show the share with any registration during the specific month. \\
\emph{Source:} Own calculation based on register data from the Swedish PES and the City of Gothenburg.}
\end{figure}

Figure \ref{fig:treatment} compares activities across treated and non-treated individuals during the 12 months after randomization. The purpose here is to show how the objectives of the program, as described in section \ref{sec:3blocks} above, match actual realizations as observed in the data. We follow the two groups both before and after the point when we did the randomization (which we denote by zero on the horizontal axis). Note again that the program activities start at the end of the first month after the randomization (i.e., May, 2017), as the beginning of the month was dedicated to preparations.

Figures \ref{fig:treatment}(a) and (b) document differences in terms of language training. We first conclude, from Figure \ref{fig:treatment}(a), that the treatment group received substantially more Swedish training at the beginning of the program period. The difference corresponds to roughly one extra week (30-- 50 hours) of training per month, which means the program adds one month (160 hours) of extra Swedish classes (green, solid line).\footnote{This increase in number of hours in language training is considerably larger that what is observed in \cite{lochmann_effect_2019} and \cite{heller_immigrant_2023}.} Next, in Figure \ref{fig:treatment}(b), we show that, in contrast to the control group, the treated individuals participate in additional, more general, courses as well. Since these courses are taught in Swedish they also add language training.

Figure \ref{fig:treatment}(c) shows what happens during the six months that follows. While there is almost no (or at least very little) supervised work practice in the control group, treated individuals exhibit a large increase just after the initial full-time classroom training has ended. We also note that the difference between treated and non-treated individuals drops to zero when the program has ended. To sum up, there is convincing evidence that the treatment group received substantially more assistance, in terms of language training and work practice, relative to the baseline. The intensified assistance starts just after the participants have entered the program, and hence constitutes an immediate injection of additional help.

The last stage of the program contains job search assistance. Since this activity is not registered in our data we cannot provide direct evidence that the treatment group received more support relative to the baseline. However, we have verified that the treatment group displays a sharp increase in the number of contacts with PES caseworkers just after the work practice ends. Since there is no such increase for the control group, we believe that the program added intensity at this final stage as well. 

\subsection{Baseline services and economic incentives} 
While we have presented a detailed description of the new integration program that we evaluate, we know less about the non-treated individuals. They get less language training, and virtually no one participates in the type of work practice offered to the treatment group. However, we want to rule out that the control group is exposed to other types of activities with a similar program intensity as the treatment group. To this end, we use PES data and identify all activities that may be defined as intensive: activities where individuals are at a workplace and/or where they get full-time training. As shown in Table \ref{tab:control} only a small fraction of the control group participates in intensive activities during the first year, which is in sharp contrast to what we observe in the treatment group. Instead, non-treated individuals end up in low-intensive activities, such as own job search or different kinds of preparatory measures.

\begin{table}[H]
  \caption{ Participation in different activities for non-treated individuals (percent).}
    \label{tab:control}
    \centering\small

\def\sym#1{\ifmmode^{#1}\else\(^{#1}\)\fi}
\begin{tabular}{l*{3}{c}}
\hline\hline
                                             &\multicolumn{1}{c}{(1)}&\multicolumn{1}{c}{(2)}&\multicolumn{1}{c}{(3)}\\
                                             &\multicolumn{1}{c}{3 months}&\multicolumn{1}{c}{6 months}&\multicolumn{1}{c}{9 months}\\
\hline
Work (including part time and\\
highly subsidized work)                                     &        7.14&       17.14&       20\\
\\
\multicolumn{3}{l}{\textbf{High-intensity activities}} \\
Training/work practice \\
(if not working)&        1.43&        2.86&        4.29\\
\\
\multicolumn{3}{l}{\textbf{Low-intensive preparatory activities }}\\
Preparatory activities \\
(if not work or high-intensive)&       45.71&       30&       31.43\\
\\
Only own job search                              &       45.71&       50&       44.29\\
\hline\hline
\end{tabular}
\floatfoot{\emph{Notes:}The category Work includes our outcome variable (having a New Start Job or having left the PES for work), part-time work and job-creation schemes. If not in some kind of work during the month, High-intensive activities include work practice and labor market training programs. If not in work or high-intensive activities, the low-intensive category includes job seekers participating in preparatory activities and job seekers only responsible for their own job search during the whole month.}

\end{table}

Newly arrived refugees that register at the Swedish PES are entitled to benefits amounting to SEK 7,000 per month. All individuals in our sample, since they are registered at the PES, receive the SEK 7,000. The individuals that participate in the integration program we evaluate are not paid any additional money --  they get the same cash transfers from the government as the individuals in the control group. Put differently, there is no direct economic incentive to participate in the more intensive activities; the only incentive is the potentially increased chance of finding a job in the end. As soon as the general integration period ends after two years the job seeker loses her right to the SEK 7,000 benefit, and, hence, there is a strong incentive for newly arrived refugees in Sweden to find a job within two years. Note that everyone in our sample faces this strong incentive to exit unemployment after two years. A job seeker who finds a job before the two-year integration program ends is not entitled to the benefit if she continues to be employed. Since the benefit is lower than a (full time) salary there is also an incentive (albeit slightly weaker) to find employment as soon as possible.\footnote{Newly arrived refugees who manage to find a job in Sweden are typically paid a monthly salary corresponding to the collectively agreed minimum wage, which is around SEK 23--24,000.} This incentive is also identical across the groups of treated and non-treated individuals.

\section{Estimation method and data} \label{sec:estimation_method_data}

\subsection{Estimation method} \label{sec:estimation_method}
When evaluating the employment effects of the new program, we estimate intention-to-treat (ITT) effects (which in the current experimental case is equal to the average treatment effect of \textit{being offered} the possibility to participate in the program). Since the procedure with summons, information meetings, and invitations to participate in labor market programs is the procedure typically used by the Swedish PES, we argue that the ITT analyses produce the most policy-relevant effect estimates.\footnote{Even though we see the ITT-estimates as the most relevant estimates, it might still be of interest for policymakers to understand if, and to what extent, \textit{actual participation} is beneficial. To examine this, we have also estimated the \textit{effects of actually participating} in the program. To obtain the average treatment effects on the treated (ATT), which in our case of one-sided compliance is equal to the Local Average Treatment Effects (LATE), we instrument actual participation with treatment status. Since the IV-estimates are given by scaling the ITT-estimates with the compliance rate, we get effects from \textit{actual} participation that are larger than the effects of \textit{being offered} to participate with a factor of $70/44$. Apart from producing larger effect estimates, the ATT-results are very similar to the ITT-results; the ATT-results are provided in the Appendix \ref{app:ATT}.} 

In this paper, we evaluate the short-run effects of the integration program. Short run here is defined as two years, and we measure the effects in each month during this follow-up period. Since we use an RCT design it is straightforward to estimate treatment effects by comparing mean outcomes across treatment and control groups, and we hence obtain the ITT-results by running a linear\footnote{This means that we rely on asymptotic properties for inference. One potential issue with this in our case is that we have an original sample of 140 individuals. Since the assignment mechanism for allocating individuals into treatment and control groups in our stratified randomized experiment is known and controlled by us, we can however also apply Fisher's approach for calculating exact p-values. Basing the inference on exact p-values instead of relying on the asymptotic properties of the estimators does however provide very similar results; see the results in the Appendix \ref{app:fischer}.} OLS-model:

\begin{align}
y_{i,t} = \alpha + \beta Treatment_i + \gamma Female_i + \epsilon_{i,t} \label{eq:OLS}
\end{align}

\noindent where $y_{i,t}$ is a discrete variable denoting whether individual $i$ was employed or not in month $t$ after randomization, $Treatment_i$ is a dummy variable indicating whether individual $i$ was assigned to the treatment group or not, and $Female_i$ is a dummy taking on the value one if individual $i$ is a woman. Note that we control for gender in these estimations since we block-randomized on gender. 

\subsection{Data}\label{sec:data}
To study the labor market effects of the integration program we use administrative data collected by the Swedish PES. The registers contain daily records on each job seeker's unemployment status, including information on open unemployment, active labor market programs, part-time and temporary work, subsidized employment, and, finally, leaving the PES for regular employment and education. We use this information to construct our outcome measure; an individual is employed in a given month if he or she is registered as either having non-subsidized work or is employed with a New Start Job (NSJ) that month.\footnote{All individuals in our sample are eligible for the subsidy NSJ, but they need to find an employer willing to hire them. With this subsidy, the employer has total wage costs of SEK 15-- 16,000 per month, which roughly corresponds to the average minimum wage level in the U.S. The subsidy is given for one year at a time for a maximum of two years and is typically used for job seekers who have been unemployed for more than 6 months.} Since we want to use a conservative measure of employment we exclude job creation schemes where employers' wage cost is very low or even zero.

The PES registers also contain information on individual characteristics, such as age, gender, education, country of origin, and date of entry to Sweden. From Table \ref{tab:balancing}, which shows descriptive statistics for the variables we have access to, it is clear that our sample consists of low-educated individuals who have recently received their residence permit: 90 percent of the individuals have no formal education above elementary school (from their country of origin), and the average number of days since the residence permit was granted is 220--230.\footnote{This means that we study an initiative that targets refugees who mainly arrived in Sweden during the refugee crisis in 2015--16. As in many other countries Sweden saw a large increase in the number of asylum seekers these years.} We also note that a majority of the individuals is men, that the average age is around 37 years, and that most of the individuals are born in Syria or in the Horn of Africa region (Eritrea and Somalia).  

Besides studying the employment effects of the integration program, we ask if, and to what extent, it affects refugees' language proficiency. For this purpose we use data provided by the the City of Gothenburg showing which course levels (A--D) each individual has completed within the SFI language training. The data also contain the number of hours in SFI training, and whether a refugee participated in the extra education mentioned in figure \ref{fig:treatment}(b).

%Since one of the main components of the integration program is intensified language training, it is relevant to

\subsubsection*{Balancing tests}
Table \ref{tab:balancing} also provides results from balancing tests. Columns one and two present the mean values and standard deviations for treatment and control groups, respectively. The third column shows differences in mean values across treated and control groups, and the final column provides p-values from tests of the null hypothesis that the mean values for the two groups are equal to zero (the tests are based on results from bivariate regressions). The p-values range from 0.184 to 0.866, indicating that the randomization managed to balance treated and non-treated individuals based on pre-determined background characteristics.

\begin{table}[ht]  \caption{Mean values and balancing on covariates, assymptotic p-values.}
    \label{tab:balancing} \footnotesize
    \centering
\begin{tabular}{l*{4}{c}}
	\hline\hline
	                               &   \multicolumn{1}{c}{(1)}   &      \multicolumn{1}{c}{(2)}    &   \multicolumn{1}{c}{(3)}   &      \multicolumn{1}{c}{(4)}  \\
	                               & \multicolumn{1}{c}{Treated} &    \multicolumn{1}{c}{Control}  &        Difference&     P-value  \\ \hline
	No formel education            &            0.171            &               0.143             &       .0286&        .645  \\
	                               &           (0.380)           &              (0.352)            &           &             \\
	Compulsory school ($\leq10$ years)      &            0.743            &               0.800             &      -.0571&        .424  \\
	                               &           (0.440)           &              (0.403)            &           &             \\
	Upper secondary school or more &           0.0857            &              0.0571             &       .0286&        .515  \\
	                               &           (0.282)           &              (0.234)            &           &             \\
	Days since residence permit    &            222.0            &               233.7             &       -11.7&        .389  \\
	                               &           (77.51)           &              (82.09)            &           &             \\
	Women                          &            0.386            &               0.371             &       .0143&        .863  \\
	                               &           (0.490)           &              (0.487)            &           &             \\
	Age                            &            37.96            &               35.70             &        2.26&        .184  \\
	                               &           (10.65)           &              (9.324)            &           &             \\
	Born in Syria                  &            0.543            &               0.557             &      -.0143&        .866  \\
	                               &           (0.502)           &              (0.500)            &           &             \\
	Born in Eritrea                &            0.157            &               0.171             &      -.0143&        .821  \\
	                               &           (0.367)           &              (0.380)            &           &             \\
	Born in Somalia                &            0.114            &              0.0714             &       .0429&        .386  \\
	                               &           (0.320)           &              (0.259)            &           &             \\
	Born in rest of the world      &            0.186            &               0.200             &      -.0143&        .832  \\
	                               &           (0.392)           &              (0.403)            &           &             \\
	Days since coming to Sweden    &            630.5            &               692.6             &       -62.1&        .442  \\
	                               &           (371.9)           &              (550.5)            &           &             \\
	Share with permanent residence &            0.443            &               0.529             &      -.0857&        .314  \\
	                               &           (0.500)           &              (0.503)            &           &             \\ \hline
	Observations                   &             70              &                70                 \\ \hline\hline
	%\multicolumn{3}{l}{\footnotesize mean coefficients; sd in parentheses}                           \\
\end{tabular}
       
\floatfoot{Notes: The table shows individual characteristics as measured at the day we did the randomization. Standard deviations in parentheses.}
\end{table}

\FloatBarrier

\subsubsection*{Refugees in Sweden: population characteristics} \label{sec:Refugees_Sweden} 
In this section we ask to what extent our sample of 140 individuals in Gothenburg are representative for the entire population of refugees that has immigrated to Sweden in the recent past years. To answer this question we use register-based micro data and compare our sample to the full population of adult refugees who got their residence permit in Sweden after 2010. From Table \ref{tab:refugee_char} it is clear that refugees in Sweden are, on average, low-educated. More than one third of those who came to Sweden as adult refugees between 2010--17 have elementary school as their highest completed degree of education. Note also that since around 20 percent of the refugees in the sample lack information on education, this figure is likely underestimated. As a comparison, it can be noted that among Swedish-born individuals, only eight percent have compulsory primary school as their highest educational degree, and among those born abroad, regardless of reason for residence permit, the corresponding figure is around 17 percent \citep{SCB2022}. It is hence clear that a large proportion of the refugees who received residence permit in Sweden during the last decade are poorly educated.

In addition, Table \ref{tab:refugee_char} shows that roughly 40 percent of adult refugees in Sweden are women, which is almost exactly the share of women (38 percent) we observe in our sample. The individuals in our study are slightly older (37) than refugees in general, but the difference is relatively small. Finally, a large majority of the refugees that have arrived in Sweden since 2010 came either from North Africa and the Middle East, or from the Horn of Africa.\footnote{In the data material we have access to, all countries of origin are grouped into larger regions. However, we know from public statistics that most refugees who came from North Africa and the Middle East have arrived from Syria during these years, that refugees from the Horn of Africa mainly come from Somalia and Eritrea, and that the refugees from areas around India are mainly from Afghanistan.} In conclusion, the characteristics presented in Table \ref{tab:refugee_char} fit well with the group of newly arrived refugees that we study in this paper.

\begin{table}[htbp]
    \centering
    \caption{Characteristics for the adult (age 18--64) refugee population in Sweden, by year of residence permit.}
    \label{tab:refugee_char}
    \begin{tabular}{l*{2}{c}}
\hline\hline
                    &\multicolumn{1}{c}{(1)}&\multicolumn{1}{c}{(2)}\\
                    &\multicolumn{1}{c}{2010--2017}&\multicolumn{1}{c}{2015--2016}\\
                    \hline
%Share refugees of all residence permit&        26.9&        41.8\\   
%\hline 
\multicolumn{3}{l}{\emph{Education at residence permit}}     \\    
$<$9 years Compulsory school&          23&        19.8\\
9-10 years of Compulsory school&        12.8&        13.9\\
Upper secondary school&        18.4&        18.8\\
Tertiary education&        25.7&        28.5\\
Information missing&        20.1&        18.9\\
\multicolumn{3}{l}{\emph{Refugee characteristics}}     \\
Women       &        42.8&        40.7\\
Mean age    &        32.5&        32.5\\  
\multicolumn{3}{l}{\emph{Region of birth}}     \\         
 North Africa and Middle East                 &        57.1&          71\\
 Horn of Africa                               &        22.1&        18.2\\
 Afghanistan, Pakistan, India, etc            &        7.36&        3.46\\
Iran/Iraq/Turkey                             &        9.13&        4.44\\
Rest of the world                            &        4.13&        2.76\\
 Rest of the world                            &        2.31&        1.47\\
 Unknown or missing                           &        .167&        .162\\   
 \hline
\(N\)                                        &      203,377&       81,053\\
 \hline\hline                
\end{tabular}
\floatfoot{\emph{Source:} Own calculations based on individual-level register data. }
\end{table}

\clearpage
%%%%%%%%%%%%%%%%%%%%%%%%%%%%%%%%%%%%%%%%%%%%%%%%%%%%%%%%%%%%%%%%%%%%%%%%%%%%%%%%
\section{Results} \label{sec:results}
%%%%%%%%%%%%%%%%%%%%%%%%%%%%%%%%%%%%%%%%%%%%%%%%%%%%%%%%%%%%%%%%%%%%%%%%%%%%%%%%
\subsection{Effects on employment}
\begin{figure}[htbp] \centering
\caption{Employment effects of the program}\label{fig:itt}
\includegraphics[width=1\textwidth]{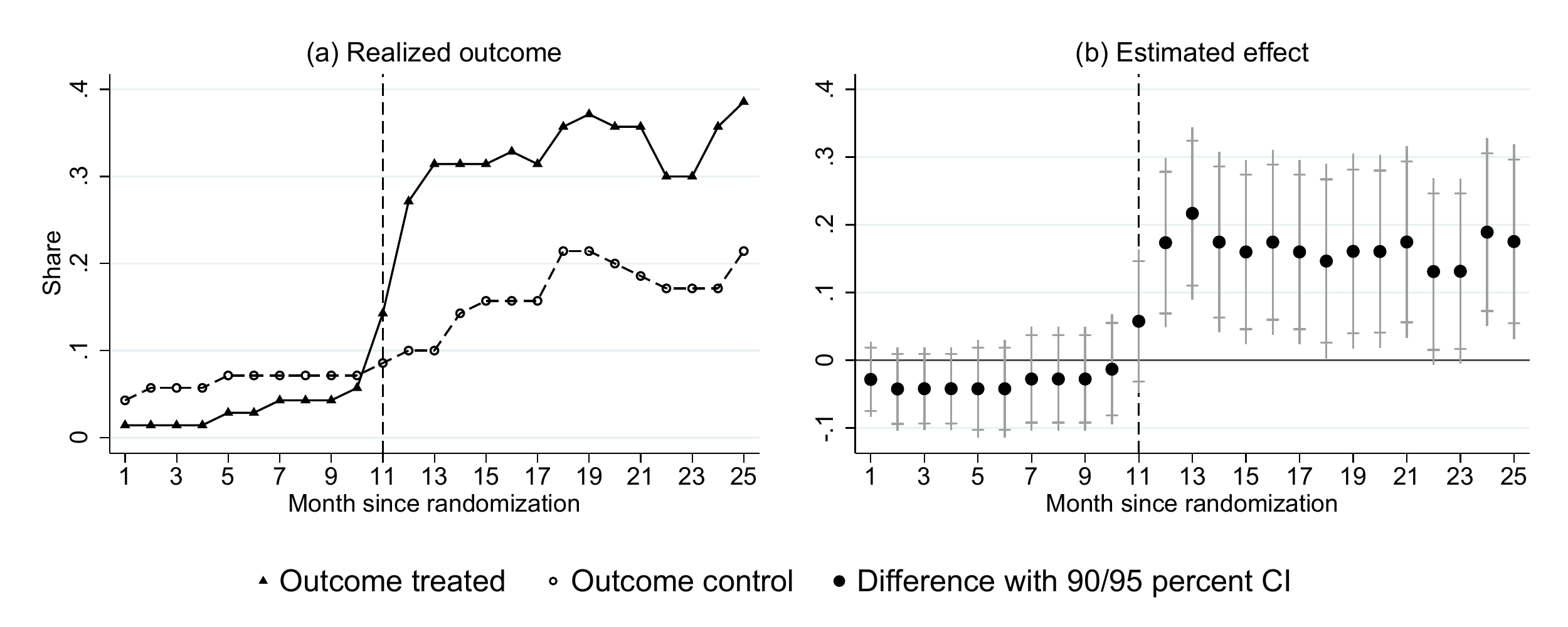} 
\floatfoot{\emph{Notes:} Figure (a) shows employment rates for treatment and control groups respectively. Figure (b) provides monthly effects on employment as given by equation (\ref{eq:OLS}) in Section \ref{sec:estimation_method}.}
\end{figure}

Figure \ref{fig:itt} illustrates the effect of the integration program on employment. We follow treated and non-treated individuals during the year when the program runs, and for one year after it has ended. We start with a simple comparison of employment levels over time in Figure \ref{fig:itt}(a). Starting with the newly arrived refugees in the control group (black circles), they exhibit a slow and gradual increase in employment rates over the two years: average employment reaches 10 percent after the first year and 20 percent at the end of the second year. This implies that the control group in our study displays similar employment growth patterns as new refugees in other OECD countries, including Norway, Denmark, Germany, and Australia. Typically, employment rates for newly arrived immigrants grows from zero to around 15--20 percent in the first 2--2,5 years after arrival (\citealp{brell_labor_2020}).

The treated individuals (black triangles) in our study display a strikingly different pattern. We first note that employment in the treatment group grows slower during the first nine months, suggesting that the program, to some extent, gives rise to lock-in effects. However, the initial period of slow growth is followed by a sharp increase in employment after ten months, at the point when the work practice ends and the job search begins. The magnitude of the increase is substantial, as employment grows from below 10 percent to above 30 percent. Comparing employment across treated and non-treated refugees suggests that the program roughly doubles the probability of getting a job, and, as seen in the figure, the effect lasts the entire second year. 

Figure \ref{fig:itt}(b) shows the corresponding monthly ITT-estimates, with 90 (caps) and 95 percent confidence intervals.\footnote{To obtain the intention-to-treat results we estimate equation (\ref{eq:OLS}) in section \ref{sec:estimation_method}. Table \ref{tab:app_p-values} in Appendix \ref{app:fischer} provides the exact point estimates and p-values from these estimations. As a sensitivity analysis, we have re-estimated the baseline model using all the variables in Table \ref{tab:balancing} as controls. The results are robust to this change, as shown in Figure \ref{fig:itt_xvar} in Appendix \ref{app:xvar}.} Taken together, there is convincing evidence of significant and substantial effects on employment. Newly arrived immigrants who get the offer to participate in the early intervention program increase their chances of being employed by 15--20 percentage points, an effect that kicks in immediately after the end of the program and is then stable over the full one-year follow-up horizon. 

\subsection{Mediation results: the role of improved language skills}
In this section, we examine how much of the effect on employment that is mediated via improved language skills. We conduct the analysis in two steps. We start by examining if the early intervention program has a positive impact on refugees' language proficiency. In the second step, we decompose the effect on employment into a component attributable to such improvements in language skills. The decomposition analysis follows \cite{Heckman2013} who examined how the Perry Preschool program in Michigan in the U.S. affected personality skills and what role these mediators had on later-in-life outcomes.\footnote{The Perry Preschool program was an influential early childhood intervention carried out in Michigan in the mid-1960s. It was implemented as an RCT where 123 children were randomly assigned to either a treatment group (58 individuals) or a control group (65 individuals), and where the studied outcomes were measured annually during ages 3--15 with additional follow-ups at ages 19, 27, and 40 \citep{Heckman2013}. \cite{Heckman2013} hence conduct their mediation analysis on data that were generated from a program with a very similar research design as in our case.}
  
Several recent studies on immigrants' labor market integration focus on the importance of language training (see, for example, \citealp{sarvimaki_integrating_2016, lochmann_effect_2019, Foged_etal_2022, Foged_Werf, heller_immigrant_2023}). Overall, the results from these studies suggest that immigrants who participate in language courses have better labor market outcomes. \cite{Foged_etal_2022} exploit a 1999 reform in Denmark and show that immigrants who are exposed to more language training have higher earnings and are more likely to be employed. \cite{sarvimaki_integrating_2016} use a similar reform in Finland and show that there is a large positive effect of language training on earnings. 

Despite the growing number of studies, there is still limited evidence on whether increased language training affects language skills, and whether increased language proficiency in turn explains any observed effects on employment.\footnote{Language training classes could, in theory, improve labor market integration by increasing refugees' social skills, network size, cultural understanding, level of job search activity etc. Hence, it is not obvious that increased language proficiency explains the positive employment effects found in the literature.} In addition, the studies that have looked closer at this question show conflicting results. While \cite{Foged_Werf} find evidence of improved language proficiency, \cite{lochmann_effect_2019} find no such effects. In this study, we use administrative records from the City of Gothenburg to capture the impact of the program on language skills. The data show the grades that recent immigrants get when they pass the different levels (A--D) within the SFI language training. The grades are set on the basis of pre-determined goals, and whether an individual meets these goals or not is decided based on teachers' assessment and the results on standardized tests.\footnote{The SFI education system is designed in the same way as the traditional education system in Sweden, with education plans that includes predetermined goals describing what the students must achieve to pass a course. There are four course levels in the SFI system, A--D. To pass a level (to get a grade), you need to pass the level, as evaluated by the teacher. You pass once you fulfill all the predetermined goals as stated in the education plan. Part of the evaluation is based on standardized tests, but the results on the tests do not unanimously determine whether a person pass or not, there is some teacher discretion.}

%Note that the only way to complete a course (level B as an example) is to pass the standardized tests at that level (the level B tests); this, in turn, means that the only way to continue to a higher level (from level B to level C, and so on) is to pass the tests. In summary, we see these grades as valid measures of refugees' language skills

\begin{figure}[htbp] \centering
\caption{Documented language skills}\label{fig:languageskills}
\includegraphics[width=1\textwidth]{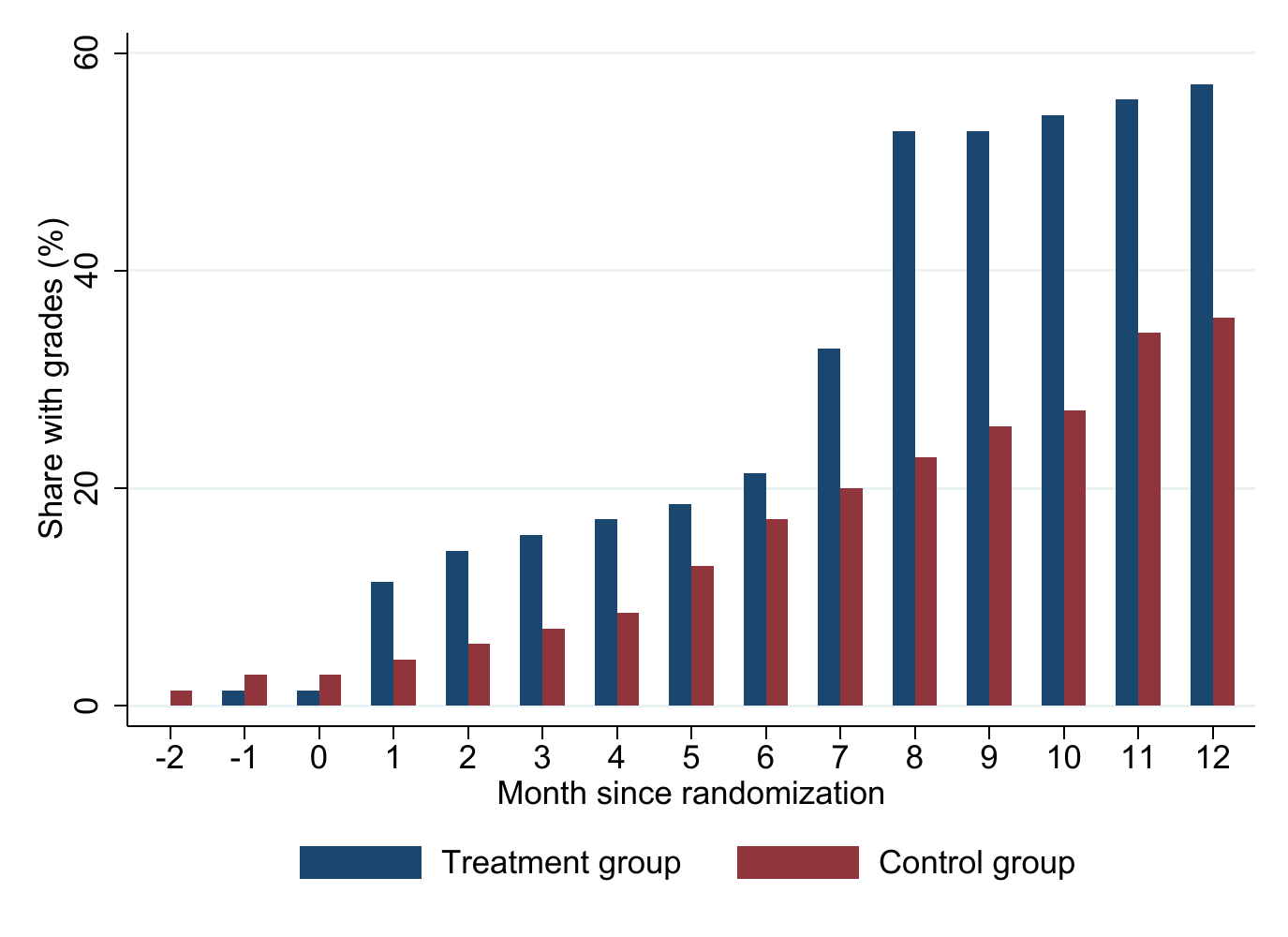} 
\end{figure}

Figure \ref{fig:languageskills} shows that language training has a strong impact on refugees’ documented language skills. The share that has a grade, and hence has completed a course within the SFI, increases sharply in the treatment group relative to the control group following the initial three months of intensive training. Table \ref{tab:language_9m} estimates the effect of the program on two measures of language skills, the probability of having any grade from SFI (column 1) and the total number of grades (column 2), both measured 9 months after randomization. Irrespective of which measure we use, there is a large and statistically significant effect of the program. The probability of having an SFI grade increases by around 27 percent, and the number of grades increases by around 0.4.

\begin{table}[ht]  \caption{Estimated program effect on language skills, nine months after randomization.}
    \label{tab:language_9m} \footnotesize
    \centering
\begin{tabular}{l*{2}{c}} \hline\hline
            &\multicolumn{1}{c}{(1)}&\multicolumn{1}{c}{(2)}\\
            &\multicolumn{1}{c}{Any grade}&\multicolumn{1}{c}{No of grades}\\
\hline
Treatment   &       0.274&       0.375\\
            &     (0.001)&     (0.001)\\
[1em]
Women       &      -0.181&      -0.233\\
            &     (0.027)&     (0.042)\\
[1em]
Constant    &       0.324&       0.401\\
            &     (0.000)&     (0.000)\\
\hline
\(N\)       &         140&         140\\
\hline\hline
\end{tabular}

\floatfoot{Notes: The estimates represent the effect of treatment on language skills as measured nine months after the randomization. p-values in parentheses.}
\end{table}

Next, we examine how much of the effect on employment that is explained by the increased Swedish language skills. The decomposition follows Heckman et al. (2013) and is conducted as follows.\footnote{This way of decomposing the effect is also in line with \cite{Gronqvist2020_JPE}.} In the first step, we estimate the following equation:

\begin{equation}\label{dec1}
   Y_i=\alpha + \beta_1 Treatment_i + \beta_2 Language_i + \beta_3 Female_i + \varepsilon_i , 
\end{equation}
where $Y_i$ is the number of months an individual is employed during the year after the program ended (months 12--25), $Treatment_i$ denotes treatment status for individual $i$, $Language_i$ measures documented language skills (any grade or number of grades), $Female_i$ is a gender dummy, $\varepsilon_i$ is the error term, and $\beta_1 - \beta_3$ are parameters to be estimated.

In the second step, we estimate the following equation:
\begin{equation}\label{dec2}
  Language_i = \gamma + \delta_1  Treatment_i +  \delta_2 Female_i + \xi_i ,
\end{equation}

\noindent where $\xi_i$ is the error term and $\delta_1 - \delta_2$ are parameters to be estimated. Using the estimated coefficients from equations (\ref{dec1}) and (\ref{dec2}) we can then calculate how large share of the estimated total effect ($\hat{\beta_1} + \hat{\beta_2} * \hat{\delta_1}$) that can be attributed to increased language skills ($\frac{\hat{\beta_2} * \hat{\delta_1}}{\hat{\beta_1} + \hat{\beta_2} * \hat{\delta_1}}$), where $\hat{\beta_2} * \hat{\delta_1}$ is the contribution from improved language skills. 

A causal interpretation of the language mediator hinges on two assumptions. First, all unobserved factors must be uncorrelated with both program participation and the mediator. Second, all unobserved factors should be orthogonal to the link between the mediator and the probability to be employed. The first assumption is likely to hold since we use an RCT design. However, randomization does not guarantee that the second assumption is fulfilled, and hence we should be careful making too strong claims about causality.

Table \ref{tab:language_dekomp} provides the decomposition results. We present results for each of the two measures of language skills. While columns (1)-(3) show the estimated coefficients, columns (4) and (5) show the calculated effect from increased language skills and the total effect of the program, and thereby finally column (6) shows the contribution from improved language skills. We find that increased language skills following program treatment explains between 6.6 and 8.6 percent of the effect on employment, depending on the measure of language skills. Note that our measures for documented language skills (see Table \ref{tab:language_9m}) may not capture all existing differences across treated and non-treated refugees in terms of language proficiency. This means that our mediation results in Table \ref{tab:language_dekomp} quite likely underestimate the importance of language skills (implying that 6.6 percent can be viewed as a lower bound). The fact that roughly 90 percent of the employment effect is still unexplained is in line with the results in Heckman et al. (2013). The residual contains all the other mechanisms that may explain the large treatment effect we find, such as improved social skills and increased networks following, e.g., supervised work practice.

\begin{table}[ht]  \caption{Decomposition of treatment effect on number of months employment, month 12--25 after randomization.}
    \label{tab:language_dekomp} \small
    \centering
\begin{tabular}{l*{7}{c}}
	\hline\hline
	             &       (1)       &       (2)       &       (3)        &               (4)                &                       (5)                        &                                           (6)                                             \\
	             & $\hat{\beta_1}$ & $\hat{\beta_2}$ & $\hat{\delta_1}$ & $\hat{\beta_2} * \hat{\delta_1}$ & $\hat{\beta_1} + \hat{\beta_2} * \hat{\delta_1}$ & $\frac{\hat{\beta_2} * \hat{\delta_1}}{\hat{\beta_1} + \hat{\beta_2} * \hat{\delta_1}}$   \\ \hline
	Any grade    &       2.2       &       .56       &       .27        &               .15                &                       2.3                        &                                          .066                                             \\
	No of grades &       2.1       &       .53       &       .37        &                .2                &                       2.3                        &                                          .086                                             \\ \hline\hline
\end{tabular} 
\end{table}

%%%%%%%%%%%%%%%%%%%%%%%%%%%%%%%%%%%%%%%%%%%%%%%%%%%%%%%%%%%%%%%%%%%%%%%%%%%%%%%%
\section{External validity}
%%%%%%%%%%%%%%%%%%%%%%%%%%%%%%%%%%%%%%%%%%%%%%%%%%%%%%%%%%%%%%%%%%%%%%%%%%%%%%%%
The integration program that we consider was designed as a direct reaction to the poor labor market situation for newly arrived refugees in the Gothenburg region. One concern is that newly arrived refugees in Gothenburg face a different environment compared to what is the case in other cities and regions in Sweden, or in other Western countries. If the Gothenburg setting is unique and very different, we may question the external validity of our study and hence the possibilities to expand the early intervention program.

\begin{figure}[htbp] \centering
\caption{Employment for similar refugees in other parts of Sweden.}\label{fig:Y_swe}
\includegraphics[width=1\textwidth]{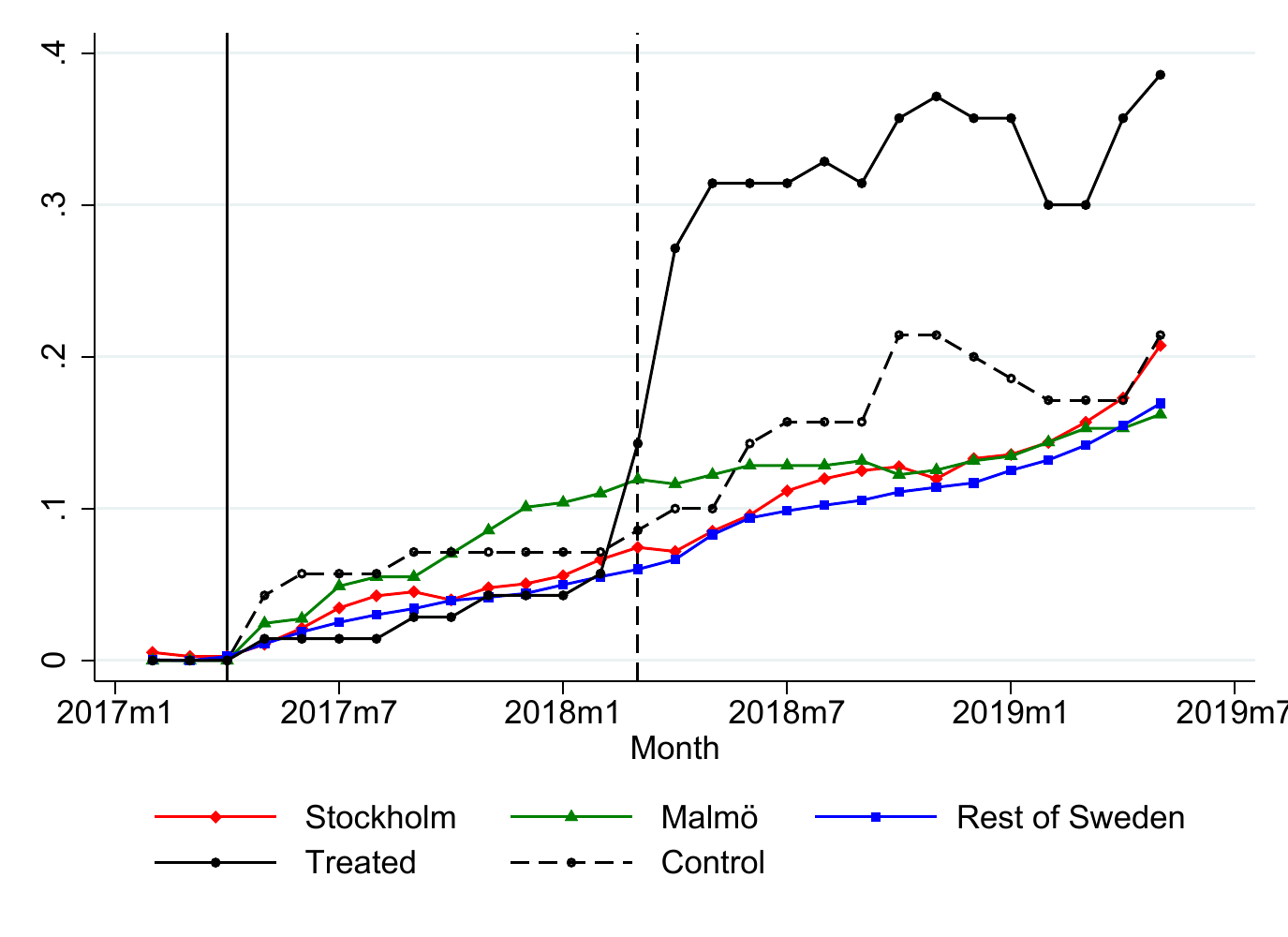} 
\floatfoot{\emph{Note:} The population consists of refugees that, by April 22 in 2017 (by the time of our randomization) (i) had at most compulsory school, (ii) were 23-64 years old, (iii) had no work disabilities, (iv) had received their residence permit within the last year, (v) had been registered at the PES at most 250 days, and (vi) were not involved in subsidized employment or in an active labor market program in April 2017. Employment is defined as having a New start job or having left the PES for work. Solid vertical line indicates the time of randomization and the vertical dashed line when the program ends.\\
\emph{Source:} Own calculations based on individual-level register data.}
\end{figure}

To address this concern we compile data for the full population of low-educated newly arrived refugees in Sweden, for years 2017--19 (see the note in Figure \ref{fig:Y_swe} for the exact selection criteria). We split the data into three groups: refugees living in Stockholm (the largest city in Sweden), refugees living in Malm\"{o} (the third largest city in Sweden), and refugees living in the rest of Sweden (excluding Stockholm, Malm\"{o}, and Gothenburg). Next, in Figure \ref{fig:Y_swe}, we compare the employment trends of these three groups to the employment trends of the treatment and the control group in our experiment.

A few things can be noted from the figure. First, the employment rates are strikingly similar across all regions just before and during the year that our program runs (the solid vertical line indicates the start of the program and the dotted vertical line shows the end of the program). This indicates that the group of refugees that we target in our study faces similar initial conditions as refugees in other parts of Sweden, hence making comparisons with other parts of Sweden reasonable. 

Second, it is clear that low-educated refugees has a hard time to establish themselves in the labor market, irrespective of where in Sweden they live. At the end of our follow up period, in May 2019 (when all refugees in the sample should have ended their introduction program at the PES) the average employment rate is at most 20 percent (in Stockholm), excluding the individuals in the treatment group. One implication of this is that the problematic labor market situation for low-educated refugees in Gothenburg, which spurred the instigation of the new program, is not unique, indicating that the need for effective integration programs is not isolated to the Gothenburg region.

Third, it is clear that the individuals in our treatment group do much better than similar refugees in other parts of Sweden. Since the individuals in our control group (receiving the baseline service at the PES) do not differ that much from similar refugees in other parts of Sweden (that also receive the baseline service at the PES), our conclusion is that refugees in other parts of Sweden also would have benefited from a program similar to the one in Gothenburg.

Finally, we note that Figure \ref{fig:Y_swe} is informative for whether displacement exist in our experiment. One concern here is that the employment effect that we observe in our study is driven by worse outcomes for the control group. That is, we may worry that the net effect on employment is smaller than what our estimates suggest -- since the treatment group may have been offered jobs that the control group would have gotten in absence of treatment. However, if displacement existed, we should expect employment in the control group to drop relative to employment in the three other groups in Figure \ref{fig:Y_swe}, since newly arrived refugees in other parts of Sweden can be assumed to be unaffected by the program in Gothenburg. That displacement seems to be a small concern in our case could be explained by the fact that the program focused on jobs where demand for labor is high and supply is low \citep{cheung_does_2023, crepon_labor_2013}.

\section{Concluding discussion}
The findings in this paper strongly support the idea that early intervention programs are successful in getting newly arrived, low-educated refugees integrated into the labor market. Our intention-to-treat estimates provide clear-cut evidence of the effectiveness of the policy as getting the offer to participate in the new integration program doubles the probability of being employed. The impact of the program is immediate, and it lasts during the whole one-year follow up period. Our study also adds new insights on the driving mechanisms behind successful labor market integration. We follow \cite{Heckman2013} and present results showing that increased host-country language skills account for 7--8 percent of the short-run effect on employment. 

Our results suggest that similar programs most likely would benefit newly arrived refugees in other cities in Sweden and in other refugee-receiving countries. Do we also think that it is practically possible to replicate our evaluated intervention in a larger scale in other countries? Evidently, the program consists of a few basic components that are highly generalizable, and most OECD countries already use these components as standard labor market policy tools \citep{card_what_2018}. Language training exists in almost all countries that accept refugee immigrants, but the key difference here is the level of intensity of the training in our program. Increasing the number of hours in such training should be possible also in other settings, as it has been done in the past in, e.g., Denmark and Finland. One important finding in our study is that early and intensive workplace training -- where the participants are expected to perform real work tasks from day one -- seem to be an effective program component. As shown by \cite{arendt_trade-offs_2023} and \cite{Arendt2022} it is however important that work practice does not crowd out language training. The strategy of using work practice as a way of helping the unemployed into employment is not unique to Sweden\citep{oecdec_how_2016}, which suggest that this component also easily replicates in other settings. 

A key factor in the program we evaluate is hence that the demand side of the labor market -- represented by the largest housing company in Gothenburg -- is involved in helping refugees acquire skills demanded by employers. The whole purpose of running the intensive work practice for six months is that the participants are trained for occupations where labor demand is high and supply is low. This situation of excess labor demand (for certain occupations) is obviously not unique to Gothenburg, implying that other Western countries, that has experienced an influx of low-educated refugees, could benefit from implementing the program as well. This conclusion is supported by evaluations of other demand driven programs (\citealp{katz_why_2022}, using an RCT design to evaluate Work Advance directed to unemployed in the US, and \citealp{foged_integrating_2022}, studying a program directed to refugees in Denmark using a staggered roll-out across municipalities to identify the effects). From a policy perspective this is a key result, as it highlights the importance of involving the demand side of the labor market to achieve faster integration. In summary, we do not see any reason why other countries, that faces challenges related to integration, would not benefit from implementing early intervention programs. 

A second central question for policy makers is how expensive the program is. We calculate the total cost of the program to be around SEK 1.2 million, corresponding to roughly SEK 27,000 per person that participated (see \ref{sec:appendix_cost}). To give the reader a sense of whether the program is expensive or not we compare our estimated cost to the cost of two other labor market programs in Sweden. We first look at the counseling program called St\"{o}d och Matchning (STOM), which is one of the most used programs for unemployed job seekers in Sweden. In this program, private providers of employment services are paid a fixed amount per participant and day, and, in addition, a lump sum if the job seeker finds employment. Our calculations show that Swedish taxpayers would have paid SEK 2.4 million only to cover the fixed cost if our treatment group instead had been offered to participate in STOM. If the STOM program has an impact comparable to the treatment effects that we find, the total cost would have increased to around SEK 2.8 million. In a second comparison, we look at the cost of labor market training programs in Sweden. The cost of such programs is calculated to be between SEK 60,000 and SEK 105,000 per participant depending on the type of training \citep{statskontoret_kostnader_2012}. Taken together, these comparisons show that the integration program that we evaluate is comparably cheap.

The immediate and positive short-run effects on employment of participating in the new integration program is very important from a policy perspective. However, even though an early attachment to the labor market in the new country is important for both refugees, and immigrants more generally, and the host country, it is also important with a stable position on the labor market in the longer run. An important task for future research is therefore to conduct a longer-run follow up of the integration program. This also makes it possible to evaluate how the importance of increased language skills plays out over a longer time horizon.     

\bibliographystyle{ecta}
\bibliography{references}

\pagebreak
\setcounter{section}{0}
\renewcommand{\thesection}{Appendix \Alph{section}}
\section{Sensitivity analyses}\label{app:sens}
\setcounter{table}{0}
\setcounter{figure}{0}
\setcounter{equation}{0}
\renewcommand{\thesubsubsection}{A.\arabic{subsubsection}}
\renewcommand{\thesubsection}{A.\arabic{subsection}}
\renewcommand{\thetable}{A\arabic{table}}
\renewcommand{\thefigure}{A\arabic{figure}}
\renewcommand{\theequation}{A\arabic{equation}}
\subsection{Inference based on Fischer's exact test}\label{app:fischer}

Since the assignment mechanism for allocating individuals into treatment and control groups in our stratified randomized experiment is known and controlled by us, we can also apply Fisher's approach for calculating exact p-values, which we do here as a sensitivity analysis. 

\subsubsection*{Fisher's sharp null hypothesis}
We adopt the null hypothesis that Fischer himself focused on, which is arguably the most obvious sharp hypothesis (\cite{imbens_rubin_2015}, p. 63). Fisher's sharp null hypothesis is given by:\footnote{Compare with the null hypothesis of no average treatment effects: $H_0: E[Y_i(1)]-E[Y_i(0)] = 0$.}
 
\begin{align}
H_0: \tau_i = Y_i(1)-Y_i(0) = 0 \qquad \forall i
\label{eq:Fischer_null}
\end{align}

\noindent The sharp null of a zero treatment effect for each individual links observed outcomes to all potential outcomes (i.e., under the null, we know both the realized and the non-realized potential outcomes). Given a test statistic, we can thus calculate its exact distribution under the null by using all possible permutations of the randomization assignments. Inference is then based on the estimated distribution of test statistics. 

\subsubsection*{Assignment mechanism}
Given the assignment mechanism for a block-randomized experiment with two strata,\footnote{The assignment mechanism for a block-randomized experiment with two strata, $m$ and $f$, is given by (\citealp{imbens_rubin_2015}, p. 191):
\begin{align}
Pr(W) &= \left(\binom{N(m)}{N_t(m)}\right)^{-1}\left(\binom{N(f)}{N_t(f)}\right)^{-1} for W \in W^+ \label{eq:assignment_mechanism}
\end{align}
\noindent where $W$ is a treatment indicator that takes on the value $1$ for observations in the treatment group and $0$ for observations in the control group, $N(m)$ is the number of males (87), $N(f)$ is the number of females (53), $N_{t}(m)$ is the number of males assigned to the treatment group (43), $N_{t}(f)$ is the number of females assigned to the treatment group (27), $W^+ = W$ such that $\sum_{i:males}W_i=N_t(m)$ and $\sum_{i:females}W_i=N_t(f)$.} we have $\binom{87}{43} \times \binom{53}{27}$ possible combinations. Since calculating a statistic for each and every one of all these possible combinations is not manageable, we will make 10,000 random draws from the set of possible combinations and base the inference on the distribution from those draws.

\subsubsection*{Test statistic}
As test statistic, we will use the absolute value of the difference in the average observed outcomes in the two strata with the relative sample sizes in each strata as weights for combining the two differences.\footnote{There is a large set of possible test statistics that can be used, but according to \cite{imbens_rubin_2015}, section 5.5, the one we use in this paper is the most popular. Formally, it is given by:

\begin{align}
\hat{\tau} &= \Big|\frac{N(m)}{N(m)+N(f)}\left(\overline{Y}_t^{obs}(m)-\overline{Y}_c^{obs}(m)\right)+\frac{N(f)}{N(m)+N(f)}\left(\overline{Y}_t^{obs}(f)-\overline{Y}_c^{obs}(f)\right) \Big| \label{eq:T_diff}
\end{align}
}
\noindent Using the assignment mechanism in equation (\ref{eq:assignment_mechanism}) to get at (random draws from) all possible combinations of the randomization assignments under the null hypothesis stated in equation (\ref{eq:Fischer_null}), we calculate the exact distribution of our statistic of interest, $F^\tau$. By comparing whether the estimated statistic in equation (\ref{eq:T_diff}), $\hat{\tau}$, is in line with $F^\tau$, we can then perform an inference based on that comparison (if few values in $F^\tau$ are higher, in absolute terms, than $\hat{\tau}$ then $\hat{\tau}$ can be considered significant).\footnote{Formally, the exact p-values are calculated as:
\begin{align}
p=\frac{1}{J} \sum_{j=1}^J1(|\Tilde{\tau_j}| \geq |\hat{\tau}|)
\end{align}
}

\subsubsection*{Results}
In Table \ref{tab:app_p-values} we show the outcome for and the difference between the treatment and the control group, together with p-values, from both inference based om asymptotic theory and from Fisher's exact test, for the 25 months following randomization. The conclusion from the table is clear, irrespective of which method for inference we use, the p-values are very similar. The differences are only large those months where we do not have any statistical significant results.

\begin{table}[ht] \caption{Asymptotic inference vs. Fisher's exact test.}
    \label{tab:app_p-values}
    \centering
    \begin{tabular}{l*{5}{c}} \hline\hline
Month since    &&&&\multicolumn{2}{c}{p-values}\\
randomization                   &     Treated&     Control&    Difference&        Asymptotic&     Fischer\\
\hline
1           &        .014&        .043&       -.028&         .32&         .62\\
2           &        .014&        .057&       -.042&         .18&         .36\\
3           &        .014&        .057&       -.042&         .18&         .37\\
4           &        .014&        .057&       -.042&         .18&         .37\\
5           &        .029&        .071&       -.042&         .25&         .45\\
6           &        .029&        .071&       -.042&         .25&         .45\\
7           &        .043&        .071&       -.028&         .48&         .72\\
8           &        .043&        .071&       -.028&         .48&         .72\\
9           &        .043&        .071&       -.028&         .48&         .72\\
10          &        .057&        .071&       -.013&         .75&           1\\
11          &         .14&        .086&        .058&         .29&          .3\\
12          &         .27&          .1&         .17&       .0072&       .0075\\
13          &         .31&          .1&         .22&       .0011&       .0013\\
14          &         .31&         .14&         .17&        .011&        .011\\
15          &         .31&         .16&         .16&        .023&        .024\\
16          &         .33&         .16&         .17&        .013&        .013\\
17          &         .31&         .16&         .16&        .023&        .022\\
18          &         .36&         .21&         .15&        .048&        .054\\
19          &         .37&         .21&         .16&         .03&        .034\\
20          &         .36&          .2&         .16&        .029&         .03\\
21          &         .36&         .19&         .17&        .017&        .018\\
22          &          .3&         .17&         .13&        .065&        .068\\
23          &          .3&         .17&         .13&        .062&        .067\\
24          &         .36&         .17&         .19&       .0084&       .0087\\
25          &         .39&         .21&         .18&        .018&         .02\\
\hline\hline
    \end{tabular}
 \floatfoot{\emph{Note:} Outcome for treatment and control group, difference, p-values, for each month since randomization. Fischer p-values from 10,000 replications.}   
\end{table}
\clearpage
\subsection{Results when controlling for covariates}\label{app:xvar}
\begin{figure}[H] \centering
\caption{Employment effects of the program, with and without controls.}\label{fig:itt_xvar}
\includegraphics[width=1\textwidth]{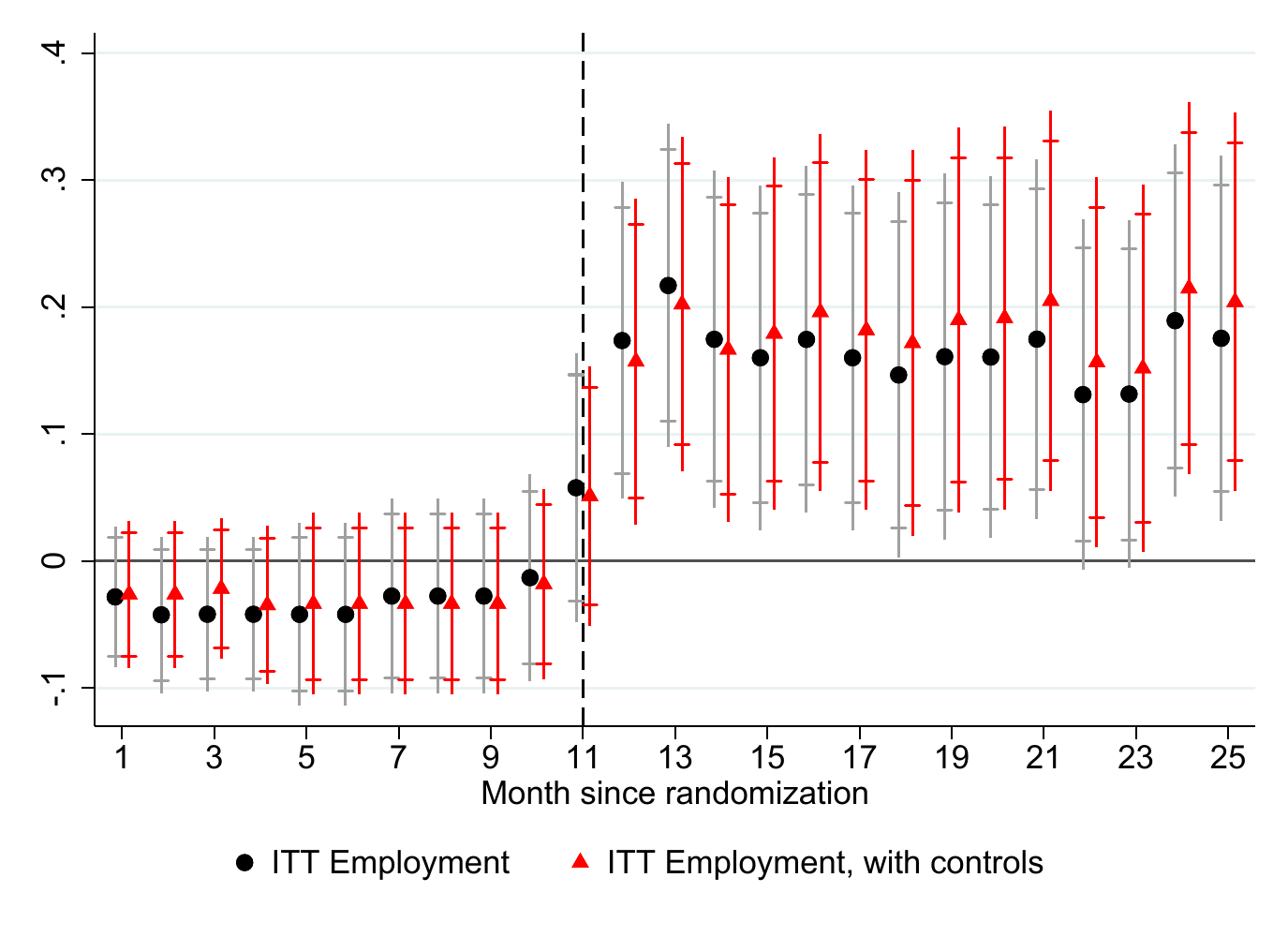} 
\floatfoot{\emph{Notes:} Black/circular dots show estimates from OLS regressions only controlling for gender (as in our main analysis), and red/triangular dots show estimates from OLS regressions controlling for all variables shown in Table \ref{tab:balancing}. Lines show 90/95 percent confidence intervals.}
\end{figure}

\clearpage
\subsection{Estimating ATT}\label{app:ATT}
\begin{figure}[htbp] \centering
\caption{Estimated effect of participation in the integration program. }\label{fig:iv}
\includegraphics[width=1\textwidth]{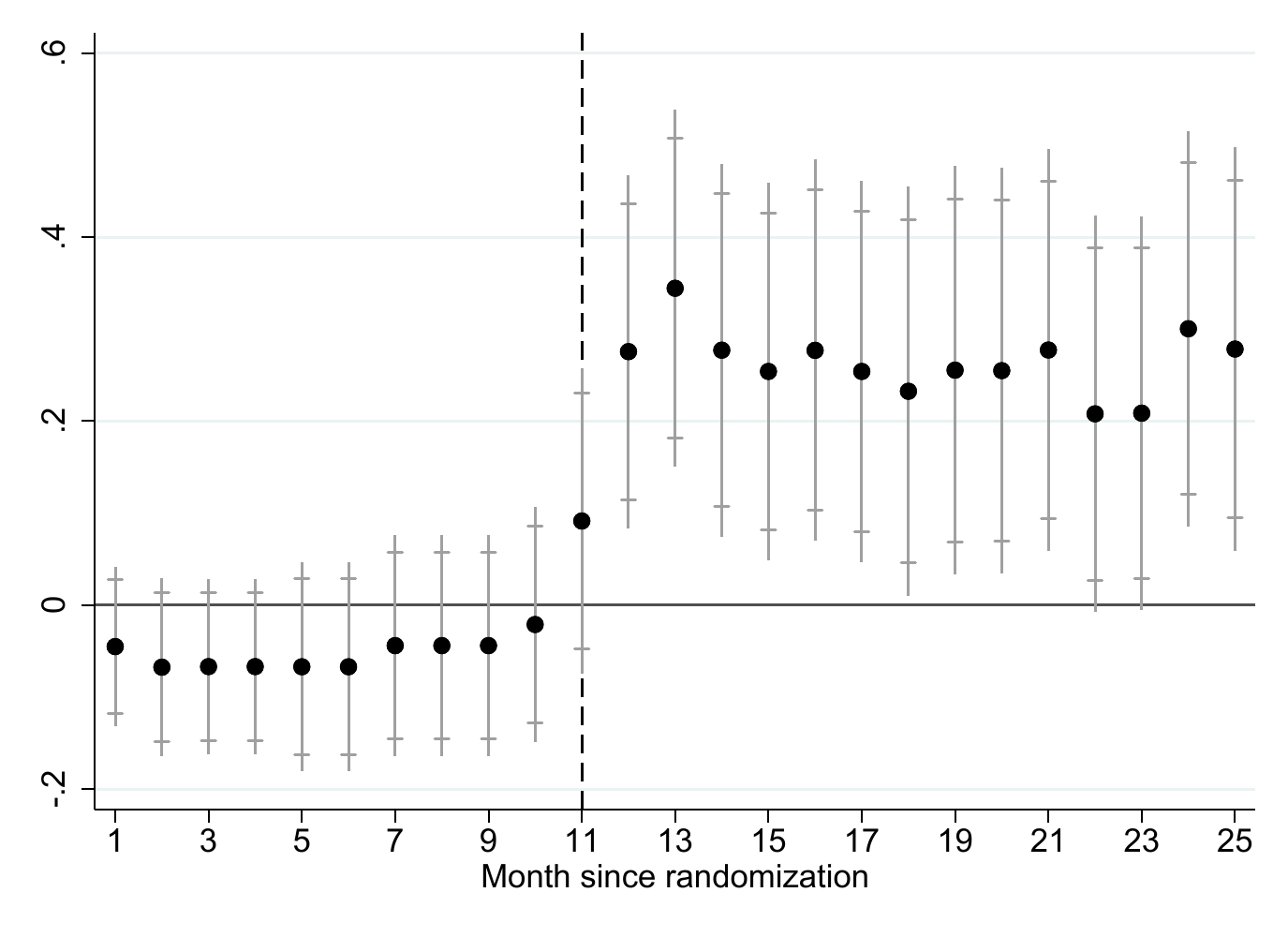} 
\floatfoot{\emph{Notes:} 2SLS estimations with 90/95 percent confidence intervals with lottery assignment as instrument for actual participation. }
\end{figure}

\section{Calculating the cost of the program \label{sec:appendix_cost}}
\setcounter{table}{0}
\renewcommand{\thetable}{B\arabic{table}}
This section presents our estimates of the cost of the integration program. The purpose is to give the reader a sense of whether the program should be considered expensive or not. Naturally, all figures that we present here should be viewed as approximations. That said, since we have tried to be conservative, we are likely overstating the true costs. 

Table \ref{tab:costs} presents our figures. We start by calculating the increased cost due to the additional teaching within the program. According to the City of Gothenburg they paid SEK 12,000 per participant, or SEK 528,000 in total, for the intensive SFI classes and other extra courses (see Figure \ref{fig:treatment}). 

Next, we consider additional costs due to the work practice. Each participant is appointed 1--3 supervisors. The supervisors, who were hand-picked from the regular staff at the workplaces, were paid SEK 50 per day for the added responsibility. While the participants typically were assigned to more than one supervisor, they were only supervised by one employee at the time. This means that the cost was a maximum of SEK 50 per day and per individual. Each participant was supervised three days per week for sixth months, corresponding to 77 days in total. Hence, the total cost was about SEK 170,000 (44 participants x 77 days x SEK 50).  

In addition, four people shared the overall operational responsibility for the work practice. According to their own approximations, they devoted on average 25 percent of full time to the program in 2016--2020. In 2017--18, which is the time window we use in our evaluation, there were two groups of program participants, our wave and the wave starting in 2016. We assume here that the time was allocated evenly across these two waves. Hence, we had one person working full time for six months with our wave, corresponding to a total wage cost of SEK 351,000.\footnote{For simplicity, we assume that the four employees have an average monthly salary of SEK 29,000, corresponding to the average monthly wage for a caseworker at the PES. The total wage cost includes payroll taxes and overhead costs.}

Finally, the program involved intensified job search assistance (JSA). The PES in Gothenburg assigned one caseworker who had the main responsibility for the extra counseling. This person worked 60 percent of full time within the program and spent the rest of the time working in the baseline services. The time spent within the program was then allocated across the different waves. In 2017--18 the caseworker devoted 20 percent of full time to our wave, corresponding to a total of 2.4 months (20 percent of full time for 12 months). In summary, this means that the PES spent around SEK 140,000 SEK in total wage costs for the intensified JSA.\footnote{We assume that the caseworker has a monthly salary of SEK 29,000 which is the average monthly wage for a caseworker at the PES. The total wage cost includes payroll taxes and overhead costs.} Note that the PES also spent money when assisting the control group in 2017--18. However, since we lack approximations of the cost of the baseline services, we just assume this cost is zero. 

In summary, additional teaching, supervised work practice, and intensified counseling costed SEK 528,000, SEK 521,000, and SEK 140,000 respectively. This adds up to SEK 1.2 million (EUR 126,000) in total, or SEK 27,000 per participant. We have also calculated the expected cost if our participants would have participated in the counseling program called St\"{o}d och Matchning (STOM). Job seekers can participate in the STOM program for a maximum of nine months. Every third month the PES decides whether a job seeker should continue or not. Since each three-month period costs SEK 18,200 per participant, the total fixed cost in our case would be SEK 2.4 million (44 individuals participate for nine months). If we assume that the STOM program has an impact that is comparable to the treatment effects that we find, roughly 50 percent of the 44 participants would have generated additional compensation to the private providers. The performance-based compensation amounts to SEK 18,000 per participant, implying the total cost amounts to around SEK 2.8 million.

\pagebreak

\begin{table}[ht!]
\begin{threeparttable}[htpb]
\caption{Program cost (SEK 1,000) during the first year after randomization.}\label{tab:costs}
\begin{tabular}{lccc}
	\hline\hline
	 & Working time (months) & Total cost\tnote{a} & Cost per participant \\ 
  \hline
  \textbf{Language training}\\
	Additional teaching\tnote{b}          &                       &           528         &        12        \\
 \\
 \textbf{Work practice}\\
	Supervisors                                 &                       &           170         &        4        \\
	Additional staff                                   &           6\tnote{c}           &           351         &        8        \\
 \\
 \textbf{Job search assistance}\\
Caseworker                    &          2.4\tnote{c}         &           140         &        3        \\
	 \hline
	Sum                                                 &                       &           1,189        &        27        \\ \hline
\end{tabular}
\begin{tablenotes}
\item [a] Wage costs are calculated using a monthly salary of SEK 29,000. We add 47.10 percent for employer contributions and pensions and 37 percent for overhead costs.
\item [b] The increased cost for additional teaching refers to more intensive SFI and some extra courses. According to the City of Gothenburg they paid SEK 12,000 per participant for these courses.
\item [c] Four people shared the operational responsibility for the work practice. Each one devoted on average 25 percent of full time to the program in 2016--2020. Since there were two waves running in parallel in 2017--2018 we had one person working full time for six months with our treatment group. 
\item [d] The figure is based on estimations made by the caseworker in Gothenburg who was responsible for the intensified counseling.
\end{tablenotes}
\end{threeparttable}
\end{table}

\end{document}